\documentclass[twocolumn,english,aps,prl,showpacs,superscriptaddress]{revtex4-1}
\usepackage[latin9]{inputenc}
\usepackage{amsmath}
\usepackage{amssymb}
\usepackage{graphicx}
\usepackage{esint}
\usepackage{physics}
\usepackage{braket}

\makeatletter

\usepackage{physics}
\usepackage{color}

\makeatother

\usepackage{babel}

\begin{document}

\title{The hybrid-order topology of weak topological insulators}

\author{Sander H. Kooi}

\affiliation{Institute for Theoretical Physics, Center for Extreme Matter and
Emergent Phenomena,\\ Utrecht University, Princetonplein 5, 3584
CC Utrecht, the Netherlands}

\author{Guido van Miert}

\affiliation{Institute for Theoretical Physics, Center for Extreme Matter and
Emergent Phenomena,\\ Utrecht University, Princetonplein 5, 3584
CC Utrecht, the Netherlands}

\author{Carmine Ortix}

\affiliation{Institute for Theoretical Physics, Center for Extreme Matter and
Emergent Phenomena,\\ Utrecht University, Princetonplein 5, 3584
CC Utrecht, the Netherlands}
\affiliation{Dipartimento di Fisica ``E. R. Caianiello'',
Universit\`a di Salerno I-84084 Fisciano (Salerno), Italy }

\date{\today} 

\begin{abstract}
We consider weak topological insulators with a twofold rotation symmetry around 
their ``dark" direction,
and show that these systems can be endowed with the topological crystalline structure of a higher-order topological insulator protected by  rotation symmetry. These hybrid-order weak topological insulators display surface Dirac cones on all surfaces. Translational symmetry breaking perturbations gap the Dirac cones on the side surfaces leaving anomalous helical hinge modes behind. We also prove that the existence of this topological phase comes about due to a 
novel
crystalline topological invariant of quantum spin-Hall insulators that can neither be revealed by symmetry indicators nor using Wilson loop invariants. Considering the minimal symmetry requirements, we anticipate that our findings could apply to a large number of weak topological insulators.
\end{abstract}

\maketitle

\paragraph{Introduction --}
The essence of a free-fermion topological insulator
is that it cannot be adiabatically deformed to a trivial atomic insulator, whose nature can be understood considering electrons as localized point particles. Put differently, topological insulators do not admit a representation in terms of exponentially localized Wannier functions (WFs). 
This obstruction to Wannier representability is, in turn, reflected in the presence of anomalous gapless boundary modes. Examples  include the chiral (helical) edge modes in quantum (spin) Hall insulators~\cite{kli80,tho82,kan05b,kan05,fu06,ber06}, as well as the surface Dirac cones of three-dimensional topological insulators (TI) \cite{fu07}.
In crystalline systems with an additional set of spatial symmetries, additional topological phases can arise~\cite{fu11}. These topological crystalline insulators (TCI) cannot be represented in terms of WFs respecting the spatial symmetries of the system, and feature, by the bulk-boundary correspondence, anomalous surface states violating a stronger version of the fermion doubling theorem~\cite{fan17} on surfaces that are left invariant under the protecting symmetry. Mirror Chern insulators~\cite{hsi12,hsi14}, for instance, are characterized by the presence of gapless surface Dirac cones pinned to mirror planes. Similarly, higher-order topological insulators (HOTI)~\cite{ben17,sch18,sch18b,mie18} feature anomalous gapless one-dimensional modes at the hinges connecting two surfaces related by the protecting crystalline symmetry~\cite{koo18}.

The topologies related to the internal and spatial symmetries do not necessarily exclude each other and can 
also
coexist. This occurs, for instance in different ``dual" topological materials~\cite{esc17,avr17,rus16,fac19}, which have  the topological structure of both a weak TI and a mirror Chern insulator. Likewise, it has been recently proposed that certain topological superconductors can concomitantly feature both surface cones and Majorana hinge modes~\cite{bul19,gho19,gho19b}. 
In all these systems, the topological crystalline structure can be diagnosed using the spatial symmetry content of the electronic bands  \cite{zak82,bra17,can18,kha18,po17} while the topology due to the internal symmetry is uniquely determined by the ``tenfold-way" invariants \cite{alt97,kit09,kru17b}. 
There exist, however, certain topological crystalline phases that are neither characterizable by symmetry indicators, nor by the tenfold way~\cite{son19fti}. In two-dimensional systems these phases have recently started to be classified~\cite{koo19}.

The question that immediately arises is whether crystalline topologies without symmetry indicators can be embedded in a topological non-trivial insulating phase protected by an internal symmetry. In this work, 
we provide an affirmative answer by 
showing that two-dimensional 
topological insulators in the wallpaper group $p2$ -- where time-reversal symmetry (TRS) guarantees the complete absence of symmetry indicators -- can be characterized by a set of three crystalline topological $\mathbb{Z}_2$ indices. They correspond to two quantized partial Berry phases~\cite{lau16,mie17} and one additional novel topological index that cannot be diagnosed even from the Wilson loop. We subsequently use this new invariant to show that weak 
TIs
possessing a twofold rotation symmetry around the $[\nu_1, \nu_2, \nu_3]$ direction, $\nu_{1,2,3}$ indicating the so-called weak topological indices~\cite{fu07b}, can be in a non-trivial topological crystalline phase. It is characterized by the presence of anomalous {\it unpinned} Dirac cones at the surfaces whose Miller indices (modulo 2) are identical to the weak topological indices, {\it i.e.} the so-called ``dark" surfaces of weak 
TIs
where  
surface Dirac cones protected by TRS
are absent. This topological crystalline phase corresponds to a form of hybrid-order topology since the system can be switched to a HOTI
with helical hinge modes using translational symmetry breaking perturbations.

\paragraph{Crystalline topological invariants in quantum spin-Hall insulators --}

\begin{figure}
 \includegraphics[width=1\columnwidth]{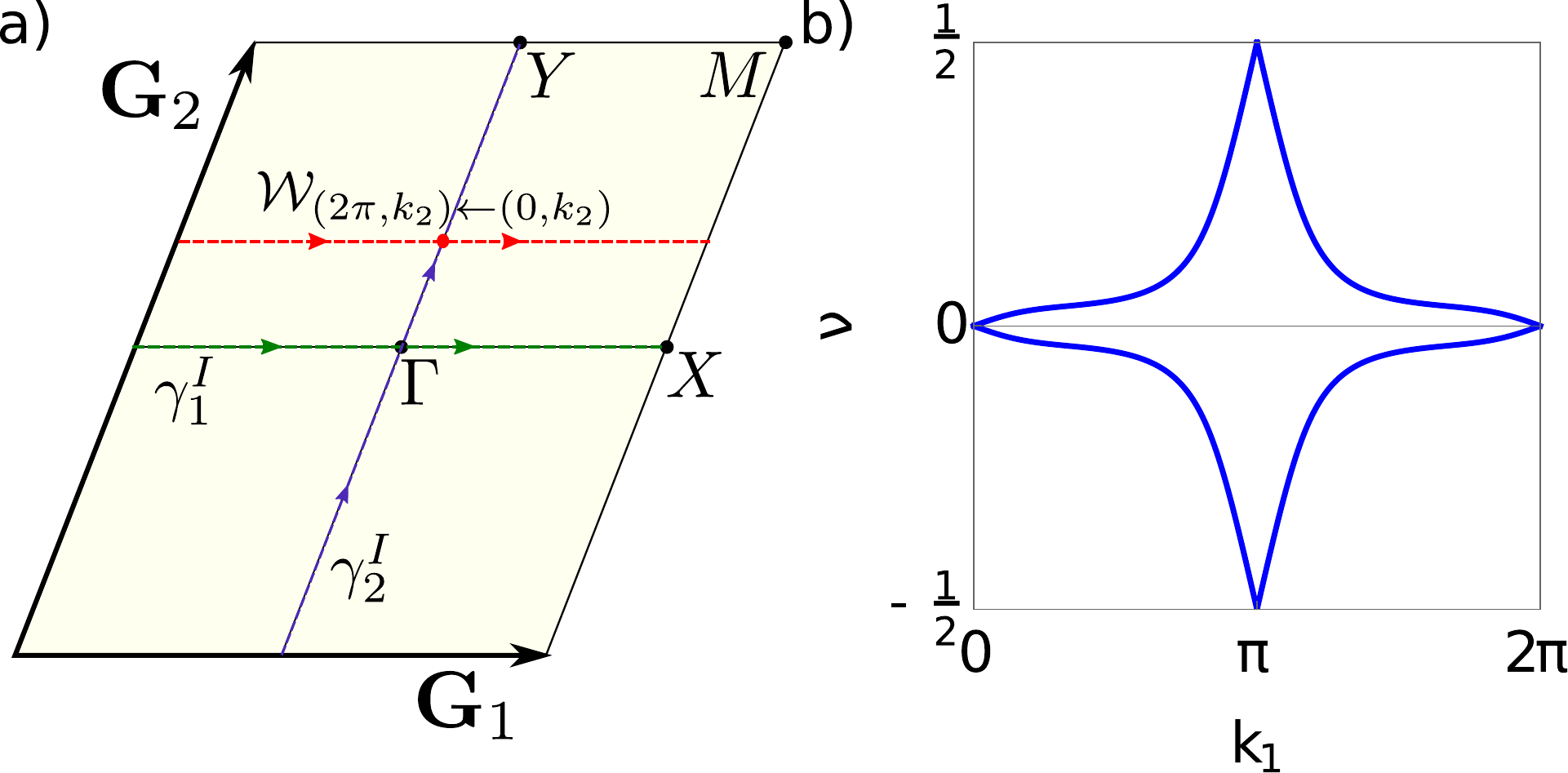}
 \caption{(a) Schematic drawing of the Brillouin zone of $\mathcal{C}_{2}$ symmetric crystal, spanned by reciprocal lattice vectors $\textbf{G}_{1}$ and $\textbf{G}_{2}$. The partial polarizations are calculated along the green and blue line, and a typical Wilson loop contour is shown in red. (b) Wilson loop spectrum of a QSHI. The winding reflects the topological non-trivial nature of the insulating phase. The quantized value of the Wilson loop for $k_{1}=0$ corresponds to the quantized partial polarizations $\gamma_2^{I}$.
 \label{fig:wilson}}
 \end{figure}

We start out by developing a
scheme 
that is able to
capture the 
full
crystalline topology of 
quantum spin-Hall insulators (QSHI) in systems with a two-fold rotation symmetry ${\mathcal C}_2$. 
To do so, we first recall that for atomic insulating phases, the crystalline topology is fully determined by the gauge-invariant charge centers~\cite{son17,koo19,fan17} of time-reversal symmetric Wannier functions that respect the symmetries of the crystal. The construction of  such symmetric Kramers pairs of Wannier functions  requires 
the construction of two time-reversed channels~\cite{fu06} of Bloch waves $\ket{\Psi_n^{I,II}(q)}$ that are separately ${\mathcal C}_2$ symmetric, where $n$ is a band index running from one to $N_F/2$ and $N_F$ the total number of occupied bands. The Bloch waves $\ket{\Psi_n^{I,II}(q)}$  need not be individual eigenstates of the Hamiltonian, but are still basis states spanning the eigenspace corresponding to the $N_F$ occupied bands. 
Importantly, the construction of symmetric Wannier functions requires a smooth, periodic, and symmetric gauge for the $\ket{\Psi_n^{I,II}(q)}$ Bloch waves.
Since we want to study crystalline topology in non Wannier representable QSHI, we relax 
these constraints on the gauge by demanding its smoothness, periodicity and symmetry modulo a $\mathcal{U}(N_F/2)\otimes \mathcal{U}(N_F/2)$ gauge degree of freedom, with these two residual gauges acting in the two time-reversed and ${\mathcal C}_2$ symmetric channels. In other words, we require a smooth, periodic and symmetric set of projectors $\rho^{I(II)}(q)=\sum_n  \ket{\Psi^{I(II)}_n(q)}\bra{\Psi^{I(II)}_n(q)}$. In the Supplemental Material we show how to construct such a gauge assuming for simplicity there are no degeneracies in the band structure other than those required by time-reversal. Since within each sector we have not demanded a continuous gauge, it follows that the channels described by the Bloch waves $\ket{\Psi_n^{I,II}(q)}$ can be characterized by non-vanishing but opposite Chern numbers $C^{I,II}$. Furthermore, the twofold rotation symmetry endows the two channels with ${\mathbb Z}$ indices that correspond to the multiplicities of the rotation eigenvalues $m_{\pm i}^{I} \equiv m_{\mp i}^{II}$ at the high-symmetry points in the Brillouin zone (BZ), {\it i.e.} $m=\Gamma,X, Y, M$ [see Fig.~\ref{fig:wilson}(a)].

We will now show that these integer crystalline indices and the Chern numbers of the channels can be used to construct four $\mathbb{Z}_2$ invariants that fully characterize the topology of $\mathcal{C}_2$ and time-reversal symmetric insulators in two dimensions. Two ${\mathbb Z}_2$ invariants 
can be immediately identified in
the quantized partial polarizations~\cite{lau16} on the ${\mathcal C}_2$ symmetric lines of the BZ $k_{1,2} \equiv 0$. 
They correspond 
to the centers of charge of one-dimensional hybrid Wannier functions and are diagnosed~\cite{koo19} by 
the Wilson loop spectra $\nu(k_{1,2})$ [see Fig.~\ref{fig:wilson}(b)]. 
These quantized partial polarizations can be expressed in terms of the crystalline indices $m_{\pm i}^{I}$ as $\gamma_{1 (2)}^{I} \equiv \left[\Gamma^I_i+X^I_i (Y^I_i) \right] \textrm{ mod } 2$ (see the Supplemental Material).
The third $\mathbb{Z}_2$ invariant corresponds to the Fu-Kane-Mele invariant that characterizes QSHI and can be expressed (see the Supplemental Material) in terms of the crystalline indices as  $\nu_{\textrm{FKM}}=\left(\Gamma^I_i+X_i^I+M_i^I+Y_i^I \right) \mod 2$. 
To define a fourth $\mathbb{Z}_2$ invariant, notice that the additional combination of eigenvalues $\nu_{1d}^{I} = \left(\Gamma^I_{-i}-X^I_{-i}-Y^I_{-i}+M^I_{-i}\right)/2 \mod 2$ is linearly independent from the previously defined $\mathbb{Z}_2$ indices. For an atomic insulating phase, $\nu_{1d}^{I}$ corresponds to the parity of the time-reversed pairs of symmetric Wannier functions centered at the corner of the unit cell with coordinates $1d= \left\{1/2, 1/2 \right\}$. The fact that in a QSHI the two time-reversed channels $I,II$ are characterized by an odd Chern number immediately yields a semi-integer value $\nu_{1d}^{I}=\pm 1/2$. However, and this is key, we can still define a $\mathbb{Z}_2$ number reading 
\begin{align*}
\gamma^I_{3}= & \dfrac{1}{2}\bigg[C^I+\left(\Gamma^I_{-i}-X^I_{-i}-Y^I_{-i}+M^I_{-i}\right)\bigg]\mod\,2.
\end{align*}
 Being independent of the partial polarizations, this new integer cannot be diagnosed by the Wilson loop spectrum but still represents a well-defined and gauge-invariant crystalline topological number. In fact,  $\gamma_3^{I}$ is manifestly gauge-invariant under intrachannel $\mathcal{U}(N_F/2)$ transformations since it is made out of a Chern number and the twofold rotation symmetry eigenvalues. Furthermore, $\gamma_3^{I}$ is also invariant under interchannel gauge transformations, which correspond to the swapping of the channels ($I \leftrightarrow II$) for isolated pairs of bands. These transformations concomitantly change the sign of the Chern numbers of the channels and the multiplicities of the ${\mathcal C}_2$ symmetry eigenvalues, and therefore do not change $\gamma_3^I$. We have thus identified three gauge-invariant ${\mathbb Z}_2$ crystalline topological indices, which together with the Fu-Kane-Mele invariant yield a ${\mathbb Z}_2^4$ classification in agreement with a recent $K$ theory study~\cite{kru19}.

We finally emphasize that the gauge-invariant $\gamma_3^I$ is different in nature from the ``spin Chern numbers" existing in systems with a mirror symmetry ${\mathcal M}_z$. In this situation, the two time-reversed and ${\mathcal C}_2$ symmetric channels $I,II$ can be taken to be the spin eigenstates $\ket{\uparrow}, \ket{\downarrow}$, such that $C^{I} \equiv C^{\uparrow}$. However, this does not determine the value of $\gamma_3^I$, as the spin Chern number does not determine $\nu^{I}_{1d}$. Thus one can find both $\gamma_3^{I}=0,1$ for the same spin Chern number.

\paragraph{Hybrid-order weak TIs -- } 
Next, we exploit the existence of 
the novel crystalline topology of $\gamma_3^I$
 in three-dimensional bulk crystals with a ${\mathcal C}_{2z}$ rotational symmetry. To do so, 
let us consider the three-dimensional Brillouin zone of our time-reversal invariant system as 
a collection of two-dimensional momentum cuts parametrized by the momentum $k_z$ parallel to the twofold rotation axis. At the time-reversal invariant two-dimensional planes $k_z=0,\pi$ we consider the system to be a topological non-trivial QSHI. 
As a result, the bulk three-dimensional crystal will be a three-dimensional topological insulator of the weak class.
In principle, we could choose the two $\mathbb{Z}_2$ topological crystalline indices corresponding to the quantized partial polarization of the $k_z=0,\pi$ QSHI to be different. This, however, would imply that in the triad of ``weak"  topological invariants~\cite{fu07b} $\left(\nu_1, \nu_2, \nu_3 \right)$,  $\nu_1$ and/or $\nu_2$ are different from zero. Hence, the three-dimensional system would feature an even number of surface Dirac cones protected by time-reversal at the $(001)$ and $(00\bar{1})$ surfaces that are left invariant under the ${\mathcal C}_{2z}$ rotation symmetry. As a result, any physical consequence of the crystalline topology cannot manifest itself: it would be completely obscured by the internal, time-reversal, symmetry topology. 

\begin{figure}
\includegraphics[width=1\columnwidth]{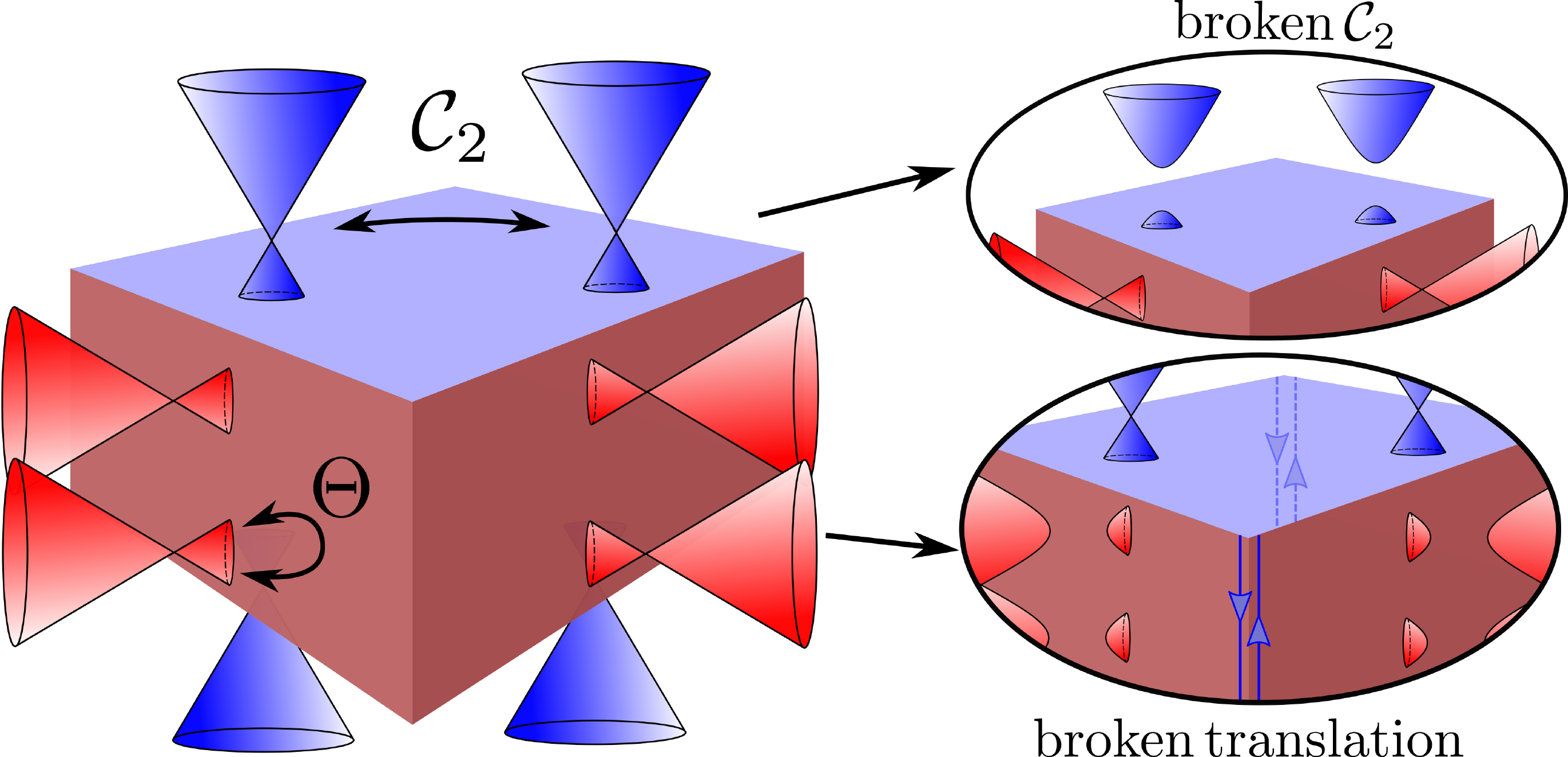}
\caption{Schematic of a hybrid-order weak topological insulator. At the top and bottom surfaces left invariant under the protecting ${\mathcal C}_{2z}$ symmetry a single pair of surface Dirac cones exist. 
On the side surfaces an even number of Dirac cones pinned to time-reversal invariant surface momenta are mandated by the weak topological invariants. 
When
breaking $\mathcal{C}_{2}$ symmetry, the topological crystalline surface Dirac cones at the top and bottom surfaces can be gapped
out leaving these surfaces completely dark. 
By breaking the translational symmetry, {\it i.e.} doubling the unit cell, the time-reversal symmetry protected Dirac cones gap out,
and the topological crystalline Dirac cones are then connected by helical hinge states.\label{fig:schematic}} 
\end{figure}

However, we can choose the two $\mathbb{Z}_2$ topological crystalline invariants at the time-reversal invariant planes to be 
equal, thus constraining the weak invariants to be $\left(0,0,1\right)$. The time-reversal symmetric topology now guarantees the existence of an even number of massless Dirac cones appearing at time-reversal invariant $(100)$ and $(010)$ surface momenta [c.f. Fig.~\ref{fig:schematic}], while the ${\mathcal C}_{2z}$ invariant $(001)$ surfaces are completely gapped. A non-trivial crystalline topology, which can thus only arise from a difference in $\gamma_3^{I}$ 
 at the $k_z=0,\pi$ planes, 
will then be in full force  and lead to the appearance of a single pair of surface Dirac cones [c.f. Fig.~\ref{fig:schematic}] at unpinned surface momenta related by the twofold rotation symmetry. This pair of surface Dirac cones realizes 
the rotational anomaly discussed in Ref.~\onlinecite{fan17}, and can be only
removed by breaking the protecting ${\mathcal C}_{2z}$ and/or $\Theta$ symmetry [c.f. Fig.~\ref{fig:schematic}]. 
We point out that the existence of this rotation anomaly 
cannot be diagnosed by considering 
the flow of gauge-invariant Wannier centers between the $k_{z}=0,\pi$ planes as in Ref.~\onlinecite{fan17}. This is because at $k_{z}=0,\pi$ our system is a topological insulator, and therefore cannot be represented in terms of localized Wannier functions. The appearance of the unpinned Dirac surface cones is instead detected by considering the $k_z$-directed Wilson loop [see the Supplemental Material] in agreement with Ref.~\cite{fid11}, although the stability of the surface Dirac cones cannot be inferred from the Wilson loop that consequently cannot be used to derive a ``topological index".
We dub this new three-dimensional insulating phase a hybrid-order weak topological insulator: it is by itself a first-order topological insulator in $d=3$ dimensions with $d-1$ gapless boundary modes, but it can be switched using unit cell doublings in the $\hat{z}$-direction, and thus without breaking any protecting symmetry, to a second-order topological crystalline insulator with anomalous gapless hinge modes  [c.f. Fig.~\ref{fig:schematic}]
and ${\mathcal C}_2$ rotation anomaly~\cite{fan17}, and reminiscent of the surface cones one predicted to appear in $\alpha-$Bi$_4$Br$_4$ and a family of Zintl compounds~\cite{hsu19,zha19tci}.

\begin{figure}
\includegraphics[width=1\columnwidth]{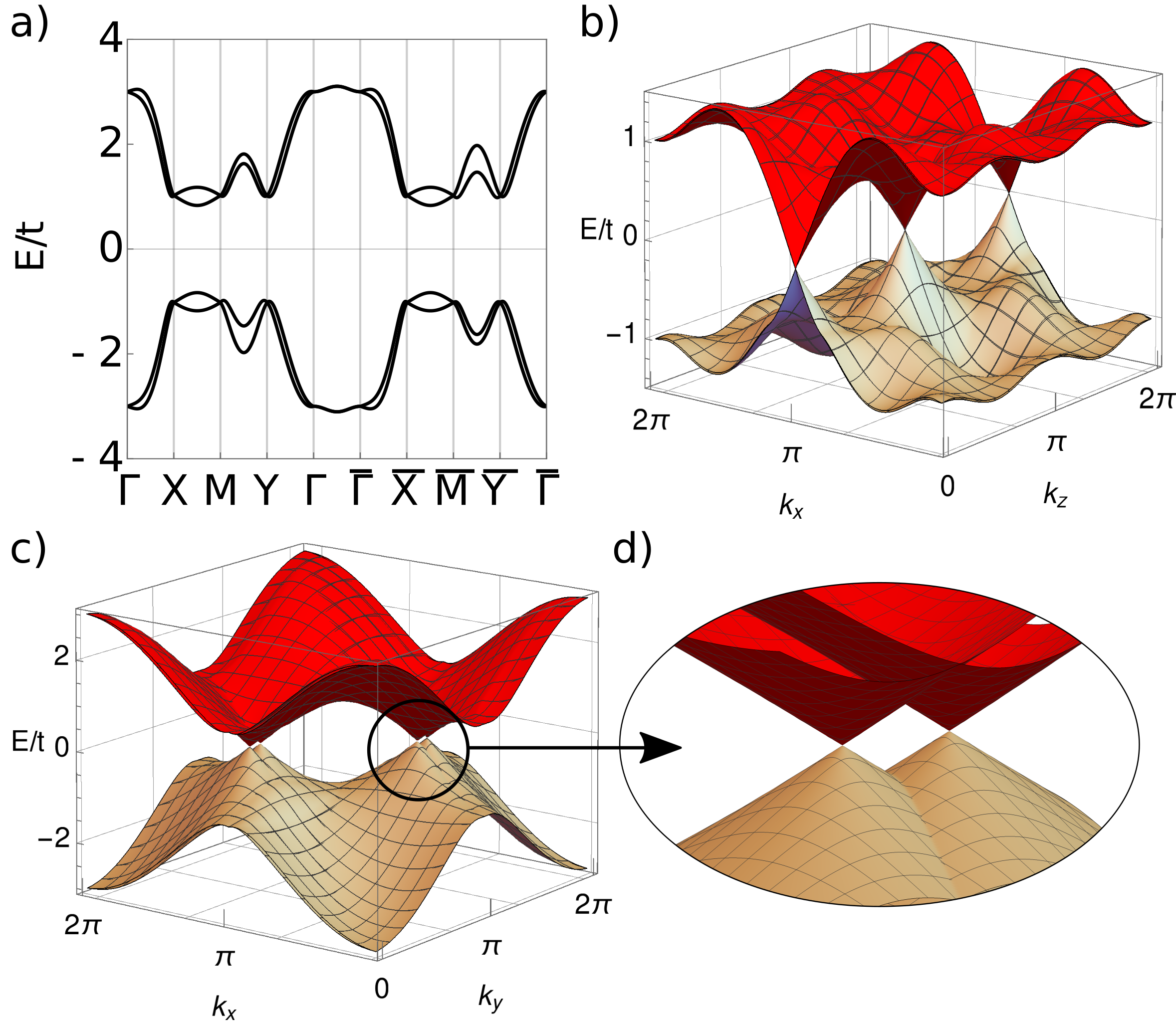}
\caption{(a) Bulk bandstructure of the stacked Kane-Mele model Eq.~\ref{eq:h1} by further accounting for a Rashba spin-orbit coupling term of strength 
$\lambda/t=0.1$. The strength of the modified intralayer spin-orbit coupling term has been fixed to $t_{2}/t=0.7$, whereas the interlayer spin-orbit coupling strength has been fixed to $t_{3}/t=0.4$. (b) Energy spectrum in a slab geometry with open boundary conditions along the $\hat{y}$ direction. The $(010)$ and $(0\bar{1}0)$ surfaces exhibit an even number of Dirac cones pinned at time-reversal invariant surface momenta as required by the weak topological invariants. (c) Surface energy spectrum along the stacking $\hat{z}$ direction. There are two pairs of surface Dirac cones localized at the $(001)$ and the $(00\bar{1})$ surface. The Dirac points are found at unpinned surface momenta related by the ${\mathcal C}_2$ symmetry. The zoom-in (d) shows that the Dirac cones at opposite surface are located at different momenta due to the lack of inversion symmetry. All energies have been measured in unit of the hopping strength $t$. 
\label{fig:1}}
\end{figure}

\paragraph{Stacked Kane-Mele model --}
Having established the existence of the hybrid-order weak topological insulator, we now present an explicit model based on stacked Kane-Mele systems realizing this phase.
 Let us consider a tight-binding model for spin-$1/2$ electrons on AA stacked honeycomb lattices. In momentum space the Bloch Hamiltonian can be written as:
\begin{align}
\mathcal{H}({\bf k}) & = d_1({\bf k}) \tau_x \otimes s_{0}+ d_2({\bf k}) \tau_y \otimes s_0 + d_5({\bf k}) \tau_{z} \otimes s_{z}\nonumber \\
 & + d_4({\bf k}) \tau_{z} \otimes s_{y},\label{eq:h1}
\end{align}
where the $\tau_i$'s and $s_i$'s are the Pauli matrices acting in sublattice and spin space respectively. The first two terms in the Hamiltonian above correspond to intralayer spin-independent nearest-neighbor hopping processes, and the corresponding coefficients are $d_1({\bf k})=-t \left[ 1 + \cos{x_1} + \cos{x_2} \right]$ and $d_2({\bf k})=- t \left[\sin{x_1} + \sin{x_2}\right]$. Here we have introduced the hopping amplitude $t$ while $x_{1,2}={\bf k} \cdot {\bf a}_{1,2}$, ${\bf a}_{1,2}$ being the Bravais lattice vectors. The third term in the Hamiltonian Eq.~\ref{eq:h1} corresponds to spin-orbit interaction which involves intralayer spin-dependent second-neighbor hopping. We take the corresponding coefficient $d_5({\bf k})= 2 t_{2}\sin\left(x_{1}\right)$, with $t_2$ the hopping strength, thus explicitly breaking the threefold rotation symmetry. Finally, the last term in the Hamiltonian involves interlayer spin-dependent hopping amplitudes and the corresponding coefficient reads $d_{4}({\bf k})=-2 t_{3} \sin\left(k_{z}\right)$. We introduce this term to explicitly break the effective ``in-plane" time-reversal symmetry~\cite{lau15}  to allow for the possibility of a change of (crystalline) topology in the two time-reversal symmetric planes $k_z=0,\pi$. Since the Hamiltonian Eq.~\ref{eq:h1} preserves bulk inversion symmetry, we can immediately obtain the strong and weak topological indices and thus obtain $\left(\nu_0 ; \nu_1, \nu_2, \nu_3 \right)=\left(0 ; 0,0,1 \right)$. In this form, however, Eq.~\ref{eq:h1} does not model a hybrid-order weak topological insulator: it can be adiabatically connected to a stack of uncoupled QSHI and consequently its $(001)$ surface 
does not feature gapless modes.
To endow the system with a non-trivial crystalline topology we instead modify the intralayer spin-orbit coupling as $d_5({\bf k}) \rightarrow \cos(k_z) d_5({\bf k})$. This modification keeps the strong and weak topological indices intact but changes the crystalline topology of the system. 
Note that also the inversion eigenvalues remain unchanged, thus implying that the hybrid-order phase cannot be diagnosed by inversion symmetry indicators.

\begin{figure}
\includegraphics[width=1\columnwidth]{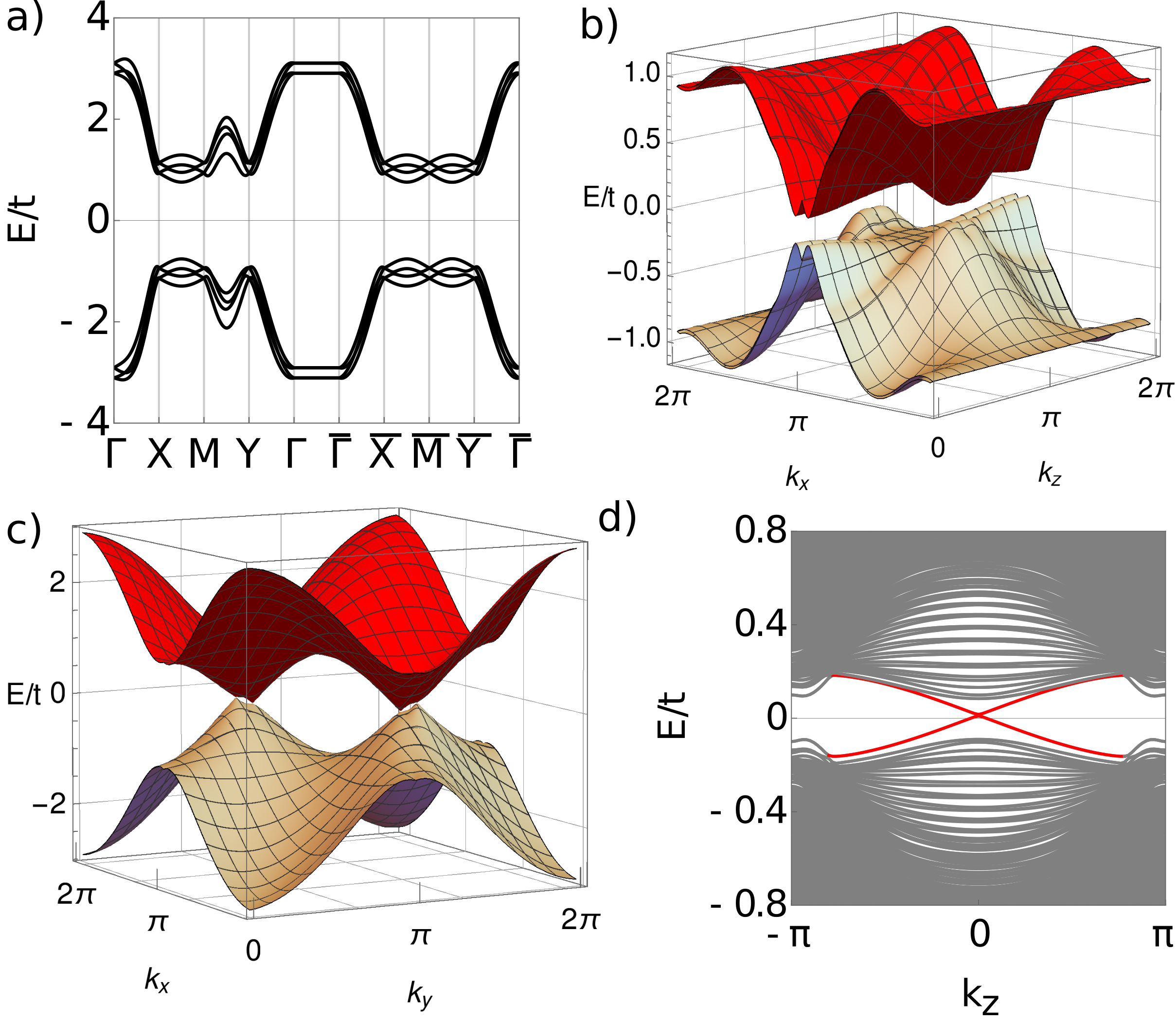}
\caption{(a) Bulk band structure of the stacked Kane-Mele model with a translational breaking perturbation. The parameter set is the same as in Fig.~\ref{fig:1}. Moreover the translational symmetry breaking parameters have been fixed to $\epsilon/t=0.1$ and $\delta/t=0.2$. (b) Surface energy spectrum showing the gapping of the time-reversal symmetry protected Dirac cones. (c) Surface energy spectrum along the stacking direction that still feature the ${\mathcal C}_2$-protected Dirac cones at unpinned surface momenta. (d) Energy spectrum in a ribbon geometry with periodic boundary conditions only along the stacking direction. Within the surface energy gap we find gapless anomalous helical hinge modes, colored in red.}
\label{fig:translation}
\end{figure}

To show this, we have computed the bulk bandstructure [see Fig.~\ref{fig:1}(a)] and the surface energy spectra [see Fig.~\ref{fig:1}(b),(c),(d)] of this modified model by further accounting for an intralayer Rashba spin-orbit coupling term \footnote{The term is $[-1/2+\cos (k_{2})-\cos (k_{1})]\tau_{y}\otimes s_{x}+ [\sin (k_{2})-\sin (k_{1})/2] \tau_{x}\otimes s_{x}+[\sqrt{3}/2 (\cos (k_{1}) - 1)] \tau_{y} \otimes \sigma_{y} + [\sin (k_{1}) \sqrt{3}/2] \tau_{x} \otimes s_{y}$} of strength $\lambda$ that explicitly breaks inversion symmetry. 
At the side surfaces we observe the conventional surface Dirac cones of a weak topological insulators [c.f. Fig.~\ref{fig:1}(b)]. More importantly, diagonalization of the Hamiltonian with open boundary conditions along the stacking direction [c.f. Fig.~\ref{fig:1}(c)] reveals the presence of two ${\mathcal C}_2$ symmetry protected surface Dirac cones thus verifying that our model realizes a hybrid-order weak topological insulator. Note that the pairs of Dirac cones at the $(001)$ and $(00\bar{1})$ surface are found at different surface momenta in agreement with the lack of inversion symmetry.

We have also verified that our model can be switched to a higher-order topological insulator by suitable translational symmetry breaking perturbations. Specifically we have introduced an interlayer staggered chemical potential of strength $\epsilon$ that provides the required doubling of the unit cell and further introduced an interlayer coupling in the enlarged unit cell of the form $-\delta\tau_{z}s_{x}$. 
Fig.~\ref{fig:translation}(a) shows the corresponding bulk bandstructure that is still characterized by a substantial gap. At the $(010)$ [$(0\bar{1}0)$] surface the time-reversal symmetry protected Dirac cones are gapped out [see Fig.~\ref{fig:translation}(b)] while the twofold rotation symmetry-protected Dirac cones at the $(001)$ [$(00\bar{1})$] surface are preserved [see Fig.~\ref{fig:translation}(c)]. Notice that Dirac cone pairs localized at opposite surface are connected by helical hinge states [see Fig.~\ref{fig:translation}(d)] as expected for a helical higher-order topological insulator protected by a twofold rotation symmetry. 

\paragraph{Conclusions --} 
To sum up, we have shown that weak topological insulators with an additional twofold rotation symmetry around the dark direction can feature a pair of Dirac cones on their dark surfaces, which are protected by the rotation symmetry. This hybrid-order weak topological insulator can be turned into a higher-order topological insulator with protected helical hinge modes by translational symmetry breaking perturbations.
We have shown that the existence of such a topological phase comes about due to a third ${\mathbb Z}_2$ topological invariant characterizing quantum spin-Hall insulators in ${\mathcal C}_2$-symmetric crystals, that can be read off neither from symmetry indicators nor from the properties of the Wilson loop spectrum. Considering the minimal symmetry requirements and the fact that the ${\mathcal C}_2$ protected surface Dirac cones appear at unpinned points in the surface Brillouin zone, we anticipate that our findings could apply to a large number of weak topological insulators.

\begin{acknowledgments}
C.O. acknowledges support from a VIDI grant (Project 680-47-543) financed by the Netherlands Organization for Scientific Research (NWO). This work is part of the research programme of the Foundation for Fundamental Research on Matter (FOM), which is part of the Netherlands Organization for Scientific Research (NWO). S.K. acknowledges support from a NWO-Graduate Program grant. 
\end{acknowledgments}

\onecolumngrid

\section*{Supplemental material for \\ "The hybrid-order topology of weak topological insulators"}

\subsection*{Constructing $\mathcal{C}_{2}$-symmetric channels for pairs of isolated
Kramers pairs}

Our goal is to numerically construct, for systems with $\mathcal{C}_{2}$
and time-reversal symmetry, a gauge that divides our system in two
channels that are by themselves $\mathcal{C}_{2}$ symmetric, and
are related to each other by time-reversal $\Theta$, and consequently $\mathcal{C}_{2}\Theta$ symmetry. 
Since we are assuming that there are no other degeneracies than those required by time-reversal symmetry, it is sufficient to consider a single Kramers pair of bands. 
We
then want to find a gauge such that we have two states $\psi_{k}^{I}$
and $\psi_{k}^{II}$ that satisfy
\begin{align}
\mathcal{C}_{2}\Theta\psi_{k}^{I} & =\psi_{k}^{II},\label{eq:cond1}\\
\mathcal{C}_{2}\psi_{k}^{I,II} & =\psi_{-k}^{I,II}.\label{eq:cond2}
\end{align}

To construct such a gauge, let us first see how to construct locally,
at each $k$-point, a gauge that satisfies Eq.~(\ref{eq:cond1}).
To do so we start by numerically diagonalizing the Hamiltonian to
find the two occupied eigenstates $\Psi_{k}^{I,II}$. We then calculate
the unitary matrix $M^{ij}=\Psi_{k}^{i\dagger}\mathcal{C}_{2}\theta\Psi_{k}^{j}$
and diagonalize it by a transformation $\text{\ensuremath{\psi_{k}^{i}=U_{k}^{ij}\Psi_{k}^{j}}},$where
$U_{k}^{ij}$ consists of the eigenvectors of $M^{ij}$. The matrix
$M$ then has the form
\begin{align*}
M & =\begin{pmatrix}e^{i\rho_{1}} & 0\\
0 & e^{i\rho_{2}}
\end{pmatrix}.
\end{align*}
 After a further unitary transformation $\psi_{k}^{1,2}\rightarrow e^{i\rho_{1,2}/2}\psi_{k}^{1,2}$,
$M$ will be the identity matrix. A final unitary transformation 
\begin{align*}
\begin{pmatrix}\psi_{k}^{I}\\
\psi_{k}^{II}
\end{pmatrix} & \rightarrow\frac{1}{\sqrt{2}}\begin{pmatrix}-1 & 1\\
1 & 1
\end{pmatrix}\begin{pmatrix}\psi_{k}^{I}\\
\psi_{k}^{II}
\end{pmatrix},
\end{align*}
then ensures $M$ is completely off-diagonal, and thus we have found
states $\psi_{k}^{I}$ and $\psi_{k}^{II}$ that satisfy Eq.~(\ref{eq:cond1}). 

In order to construct a gauge satisfying Eqs.~(\ref{eq:cond1}) and
(\ref{eq:cond2}) across the Brillouin zone, let us start by noting
that for 
an isolated Kramers pair of bands, 
the local gauge freedom is $U(1)\otimes SU(2)$.
That is, we are free to do a rotation between the basis states, and
can then change each basis state by a $U(1)$ phase. Now the conditions
of Eqs.~(\ref{eq:cond1}) and \eqref{eq:cond2} completely fix the $SU(2)$ part of the
gauge freedom, up to the unitary operation of relabeling the states $\psi_{k}^{I}\longleftrightarrow\psi_{k}^{II}$. There will thus be a remaining $U(1)$ degree of freedom within each $\mathcal{C}_2$ symmetric channel.

We can now start at the $\Gamma$ point in the Brillouin zone at $k=(0,0)$,
and follow the steps described to make the matrix $M$ off-diagonal.
Since $\Gamma$ is a time-reversal invariant point, Kramers theorem
guarentees that the time-reversal operator $\theta$ is off-diagonal. Rotating such that $C_2 \theta$ is off-diagonal thus also ensures that this basis is $C_2$ diagonal.

Once we have found the states at the $\Gamma$ point, we take a point
close to it at $k=(\epsilon,0)$ for a small $\epsilon$ and repeat
the procedure to make $M$ off-diagonal. We then consider the overlap
matrix $\psi_{(0,0)}^{i}\psi_{(\epsilon,0)}^{j}$, which will either
be almost completely diagonal, or almost completely off-diagonal.
If it is off-diagonal, this means that we need to switch the labels
$I,II$ by a unitary transformation
\begin{align*}
\begin{pmatrix}\psi_{k}^{I}\\
\psi_{k}^{II}
\end{pmatrix} & \rightarrow\begin{pmatrix}0 & 1\\
1 & 0
\end{pmatrix}\begin{pmatrix}\psi_{k}^{I}\\
\psi_{k}^{II}
\end{pmatrix}.
\end{align*}
We then define $\psi_{-k}^{i}=\mathcal{C}_{2}\psi_{k}^{i}$. Continuing
in this way along half the Brillouin zone also ensures that the second
condition Eq.~(\ref{eq:cond1}) is satisfied.

The gauge constructed in this way fixes the $SU(2)$ part of the gauge degrees of freedom, since we ensured that the overlap matrix is diagonal. Within each channel, there is still a $U(1)$ phase degree of freedom which we have not fixed, and for which there is an obstruction to smoothness if the band has a non-zero Chern number. For our purposes we do not need to fix this phase, as neither the eigenvalues nor the Chern number can change under this $U(1)$ gauge transformation

\subsection*{Topological invariants of systems with two-fold rotational symmetry}

Using the procedure of the previous section, we can divide the occupied bands of a system into two channels that are tow-fold rotation symmetric, and have broken time-reversal symmetry. In this section we show how one can compute the topology of the system once such a gauge has been found.

First consider the partial polarizations, which are quantized along the high-symmetry lines of the BZ by $\mathcal{C}_2$ and time-reversal symmetry. Using the constructed gauge, the two $\mathcal{C}_2$-symmetric channels give us an easy way to compute the partial polarizations. Since the two channels are related by time-reversal symmetry, the partial polarization is equal to the polarization in one channel. This means we have to calculate the Berry phase of one channel along high-symmetry lines in the BZ.
On these contours, each channel is effectively a one-dimensional system with $\mathcal{C}_2$ symmetry. For such systems, the Berry phase $\gamma$ is quantized and can directly be expressed in terms of its $\mathcal{C}_2$-eigenvalues as \cite{hug11}

\begin{align*}
\gamma   = & (-i) \prod_i^{N_{F}}\log[-\xi_i(0) \xi_i(\pi)] \mod 2 \pi,  
\end{align*}

where $\xi_i(k)$ is the $\mathcal{C}_2$ eigenvalue of band $i$ at momentum $k$. Since the eigenvalues can be either $\pm i$, the term inside the logarithm can either be $\pm$ and hence each term inside the product is either $0$ or $\pi$. This allows us to rewrite

\begin{align*}
\gamma  = & \sum_i^{N_{F}} -i \log[-\xi_i(0) \xi_i(\pi)] \mod 2 \pi \\  
    = & \sum_i^{N_{F}} -i (\log[i\xi_i(0)]+\log[i\xi_i(\pi)]) \mod 2 \pi \\
    = & -i\sum_i^{N_{F}}\log[i\xi_i(0)]-i \sum_i^{N_{F}}\log[i\xi_i(\pi)]) \mod 2 \pi \\
    = & \pi (m^{k=0}_i+m^{k=\pi}_{i}) \mod 2 \pi,
\end{align*}

Where $m^{k=0}_i$ denotes the multiplicity of eigenvalue $i$ at $k=0$ and the last equality follows since whenever an eigenvalue is $-i$, the log evaluates to zero.

Applying this to the contours $\Gamma-X$ and $\Gamma-Y$ we find $\gamma_{1 (2)}^{I} \equiv \left[\Gamma^I_i+X^I_i (Y^I_i) \right] \textrm{ mod } 2$ as presented in the main text.

The Fu-Kane-Mele invariant is equal to the Chern number in one of the channels modulo 2. This Chern number can in turn be expressed as the difference of the berry phases of the contours $\Gamma-Y$ and $X-M$ \cite{kru19}. Hence the FKM-invariant can be expressed as

\begin{align*}
\nu_{\mathrm{FKM}}  = & (\Gamma^I_i + X^I_i + Y^I_i + M^I_i) \mod 2  \\   
\end{align*}

So far we have found three invariants constructed out of four eigenvalue multiplicities. For atomic insulators, a fourth linearly independent crystalline invariant is given by \cite{mie18}

\begin{align*}
\nu_{1d} = & \frac{1}{2}(\Gamma_{-i}-X_{-i}-Y_{-i}+M_{-i}),   
\end{align*}
where $m_{-i}$ is the multiplicity of eigenvalue $-i$ at high-symmetry point $m=\Gamma,X,Y,M$.

This invariant is related to the number of Wannier functions localized at Wyckoff position $1d$ in real space by \cite{mie18}

\begin{align*}
\nu_{1d} = & -N_{1d;i}+N_{1d;-i}
\end{align*}

where $N_{1d;i}$ is the number of symmetric Wannier functions localized at Wyckoff position $1d$ with two-fold rotation eigenvalue $i$. When considering non-atomic insulators, one cannot find localized symmetric Wannier functions, however we can still calculate $\nu_{1d}$. For systems with an odd Chern number $\nu_{1d}$ will be half-integer, since there will be an odd number of $i$ eigenvalues, and an odd number of $-i$ eigenvalues. Hence we can construct an invariant that is always integer by the combination

\begin{align*}
\gamma_3 = (\mathcal{C}/2+\nu_{1d}) \mod 2.
\end{align*}

\subsection*{Sums of quantum spin Hall insulators}

Here we explicitly show how adding together and coupling two QSHI
can result in either a fragile topological phase or a trivial phase.
To this end we consider a Kane-Mele model, 
\begin{align*}
\mathcal{H}_{KM}\left(t,t_{2}\right) & =-t\left[1+\cos\left(k_{1}\right)+\cos\left(k_{2}\right)\right]\tau_{x}\otimes s_{0}\\
 & -t\left[\sin\left(k_{1}\right)+\sin\left(k_{2}\right)\right]\tau_{y}\otimes s_{0}+2t_{2}\sin\left(k_{1}\right)\tau_{z}\otimes s_{z},
\end{align*}
where $t$ and $t_{2}$ are the nearest-neighbor and intrinsic spin-orbit
coupling amplitudes, and $\tau_{i}$ and $s_{i}$ are Pauli matrices
acting in sub-lattice and spin-space. We also add a Rashba coupling
\begin{align*}
\mathcal{H}_{R}= & \lambda \Big\{ [-1/2+\cos (k_{2})-\cos (k_{1})]\tau_{y}\otimes s_{x}+ [\sin (k_{2})-\sin (k_{1})/2] \tau_{x}\otimes s_{x}+[\sqrt{3}/2 (\cos (k_{1}) - 1)] \tau_{y} \otimes \sigma_{y} \\
+ & [\sin (k_{1}) \sqrt{3}/2] \tau_{x} \otimes s_{y}\Big\},
\end{align*}
where $\lambda$ is the Rashba amplitude. Note that the intrinsic
spin-orbit coupling term only acts along $k_{1}$ such that threefold
rotation symmetry is broken.

\begin{figure}
\includegraphics[width=\textwidth]{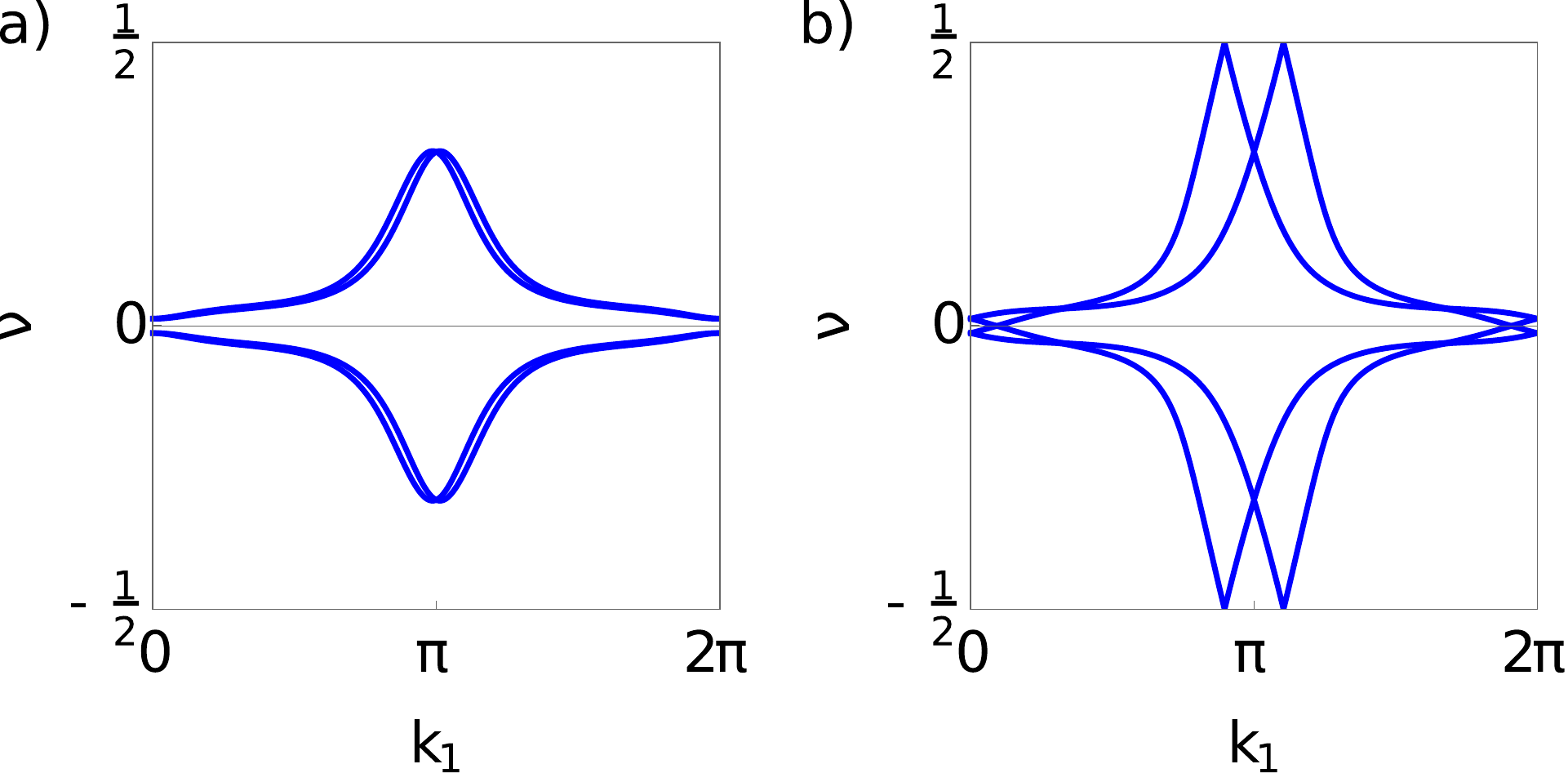}
\caption{(a) Wilson loop spectrum of two coupled Kane-Mele models for $t^{2}=t^{1}$, $t^{1}_{2}/t^{1}=t^{2}_{2}/t^{1}=0.7$, $\lambda/t^{1}=0.2$ and $\delta/t^{1}=0.4$. (b) Wilson loop spectrum for the same model as in a), changing only $t^{1}_{2}/t^{1}=-t^{2}_{2}/t^{1}=0.7$.}
\label{fig:wlsKM}
\end{figure}

We now take two copies of this model, and couple them by adding a
term $-\delta i\sigma_{x}\otimes\tau_{x}\otimes s_{y}$, where $\delta$
is the amplitude of the coupling and $\sigma_{x}$ is a Pauli matrix
acting in \char`\"{}copy\char`\"{} space. Let us denote the parameters
of the two copies by $t^{i}$ and $t_{2}^{i}$ , where $i=1,2$ denotes
the copy. To examine the phase that results from coupling the two QSHI, we consider
its Wilson loop spectrum along $k_{1}$ (see main text). Let us first
consider the parameters of the two Kane-Mele models to be equal and
such that the system is insulating. This leads to a Wilson loop spectrum as in Fig.~\ref{fig:wlsKM}(a), showing
this is an atomic insulator with two Kramers pairs at unpinned points
in the BZ. Changing the sign of $t_{2}$ between the copies, such
that $t_{2}^{1}=-t_{2}^{2}$ and keeping the sign of $t$ the same,
results in a fragile topological phase {[}see Fig.~\ref{fig:wlsKM}(b){]}. Note that when turning off Rashba coupling, such that we restore inversion symmetry, the two copies have the same inversion eigenvalues, meaning this phase or $\gamma_3$ cannot be diagnosed by inversion. To obtain
this phase, we can also instead change the sign of $t$ between the
copies such that $t^{1}=-t^{2}$, while keeping the sign of the other
parameters equal. Changing both the sign of $t$ and $t_{2}$ between
the two copies again results in the trivial phase.

Finally, we note that when taking the hybrid-order topology model and adding translational-symmetry breaking perturbation (as presented in the main text), the resulting system at $k_z=0$ will be fragile topological insulator with Wilson loop Fig.~\ref{fig:wlsKM}(b), while at $k_z=\pi$ it will be a trivial insulator with Wilson loop Fig.~\ref{fig:wlsKM}(a).

\subsection*{Spectral flow in the hybrid-order topological insulator}

\begin{figure}[h!]
\includegraphics[width=0.8\columnwidth]{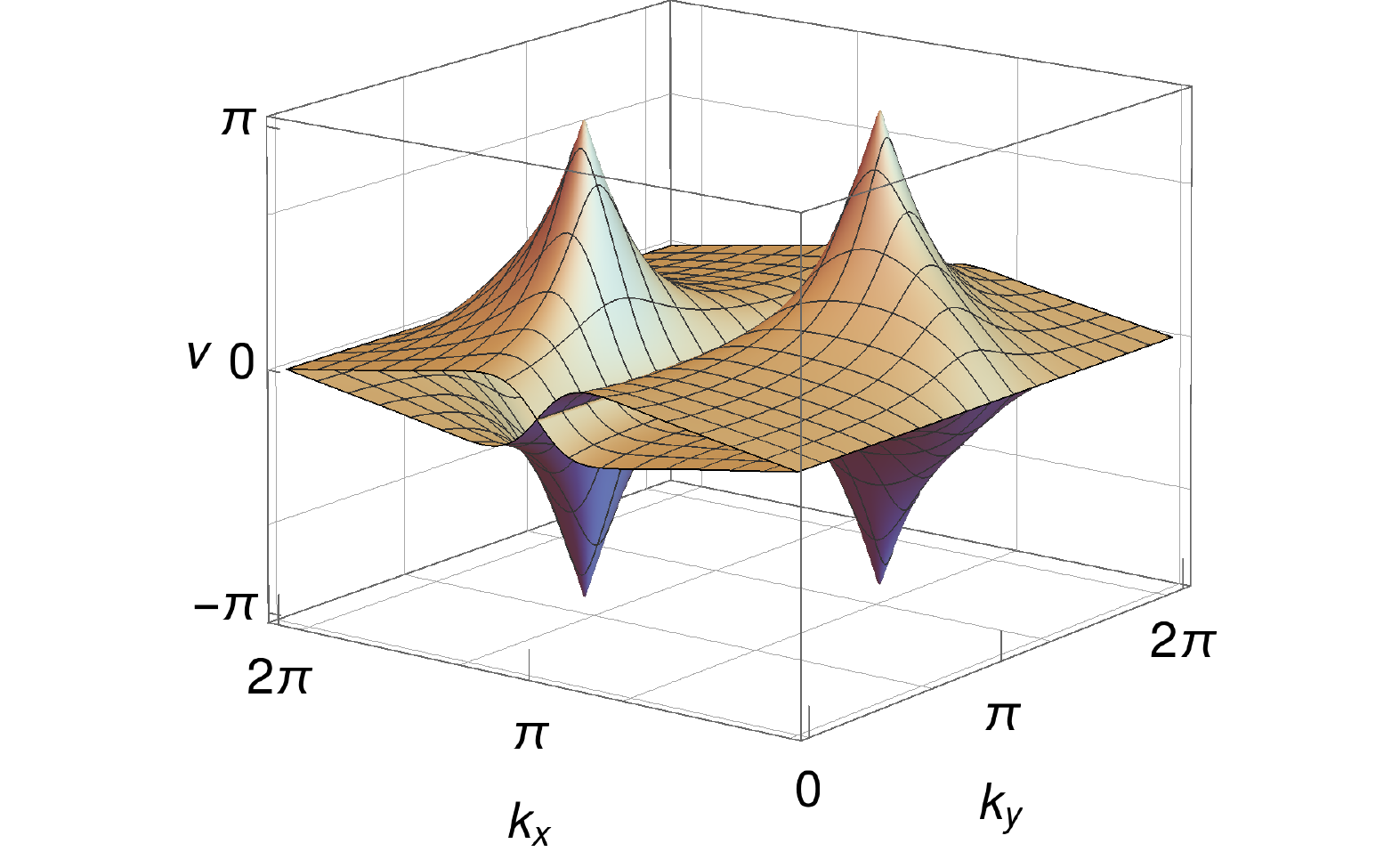}
\caption{Wilson loop spectrum along $k_{z}$ plotted as a function of $k_{1}$ and $k_{2}$ of the hybrid-order topological insulator presented in the main text. Two $\mathcal{C}_{2}$ related band crossings indicate gapless modes on the top and bottom surfaces.}
\label{fig:wlsKM3d}
\end{figure}

In general protected edge states can be detected by studying the Wilson loop spectrum along the momentum direction perpendicular to the surface \cite{fid11}. For the hybrid-order topological insulator, the $\mathcal{C}_{2}$-related cones on the top and bottom surfaces can be detected by considering the Wilson loop along $k_{z}$, as a function of $k_{1}$ and $k_{2}$. This is plotted in Fig.~\ref{fig:wlsKM3d} for the hybrid-order topological insulator presented in the main text. Here we see two protected gap-closing points that are related by $\mathcal{C}_{2}$ symmetry.

\subsection*{Channel decomposition of the  hybrid-order topological insulator}
In Table \ref{tab:1} provide the multiplicities of the ${\mathcal C}_2$ eigenvalues in the two time-reversed and rotation-symmetric  channels for our pristine hybrid order topological phase. Then we consider a dimerization in the $z$-direction that trivializes the time-reversal invariant weak topological indices but does not change the value of $\gamma_3^I$ (Table \ref{tab:2}). Finally, in Table \ref{tab:3} we consider a dimerization in the $x$-direction, which does not change the weak topological indices, nor the $\mathbb{Z}_2$ index governing the appearance of unpinned Dirac cones at the bottom and top surfaces given by $[\gamma_3^I(k_z=\pi) - \gamma_3^I (k_z=0)]\mod 2$.

\begin{table}
\centering
\begin{tabular}{|c|c|c|c|c|c|}
\hline 
 & $C^{I}$ & $\Gamma^{I}$ & $X^{I}$ & $Y^{I}$ & $M^{I}$\tabularnewline
\hline 
$k_{z}=0$ & 1 & $+i$ & $+i$ & $+i$ & $-i$\tabularnewline
\hline 
$k_{z}=\pi$ & 1 & $-i$ & $-i$ & $-i$ & $+i$\tabularnewline
\hline 
\end{tabular}

\begin{tabular}{|c|c|c|c|c|c|}
\hline 
 & $C^{II}$ & $\Gamma^{II}$ & $X^{II}$ & $Y^{II}$ & $M^{II}$\tabularnewline
\hline 
$k_{z}=0$ & -1 & $-i$ & $-i$ & $-i$ & $+i$\tabularnewline
\hline 
$k_{z}=\pi$ & -1 & $+i$ & $+i$ & $+i$ & $-i$\tabularnewline
\hline 
\end{tabular}

\caption{Tables showing the channel eigenvalues and channel Chern number of
the two channels $I$ and $II$ at $k_{z}=0,\pi$ of the hybrid-order topological insulator.\label{tab:1}}

\end{table}

\begin{table}
\begin{tabular}{|c|c|c|c|c|c|}
\hline 
 & $C^{I}$ & $\Gamma^{I}$ & $X^{I}$ & $Y^{I}$ & $M^{I}$\tabularnewline
\hline 
$k_{z}=0$ & $2$ & $\{-i,+i\}$ & $\{-i,+i\}$ & $\{-i,+i\}$ & $\{-i,+i\}$\tabularnewline
\hline 
$k_{z}=\pi$ & $0$ & $\{+i,+i\}$ & $\{+i,+i\}$ & $\{+i,+i\}$ & $\{-i,-i\}$\tabularnewline
\hline 
\end{tabular}

\begin{tabular}{|c|c|c|c|c|c|}
\hline 
 & $C^{II}$ & $\Gamma^{II}$ & $X^{II}$ & $Y^{II}$ & $M^{II}$\tabularnewline
\hline 
$k_{z}=0$ & $-2$ & $\{-i,+i\}$ & $\{-i,+i\}$ & $\{-i,+i\}$ & $\{-i,+i\}$\tabularnewline
\hline 
$k_{z}=\pi$ & $0$ & $\{-i,-i\}$ & $\{-i,-i\}$ & $\{-i,-i\}$ & $\{+i,+i\}$\tabularnewline
\hline 
\end{tabular}

\caption{Tables showing the channel eigenvalues and channel Chern number of
the two channels $I$ and $II$ for the hybrid-order topological insulator
with a translational-symmetry perturbation in the $z$-direction at
$k_{z}=0,\pi$.\label{tab:2}}
\end{table}

\begin{table}
\begin{tabular}{|c|c|c|c|c|c|}
\hline 
 & $C^{I}$ & $\Gamma^{I}$ & $X^{I}$ & $Y^{I}$ & $M^{I}$\tabularnewline
\hline 
$k_{z}=0$ & $1$ & $\{+i,+i\}$ & $\{-i,+i\}$ & $\{-i,+i\}$ & $\{-i,+i\}$\tabularnewline
\hline 
$k_{z}=\pi$ & $1$ & $\{-i,-i\}$ & $\{+i,+i\}$ & $\{+i,+i\}$ & $\{-i,-i\}$\tabularnewline
\hline 
\end{tabular}

\begin{tabular}{|c|c|c|c|c|c|}
\hline 
 & $C^{II}$ & $\Gamma^{II}$ & $X^{II}$ & $Y^{II}$ & $M^{II}$\tabularnewline
\hline 
$k_{z}=0$ & $-1$ & $\{-i,-i\}$ & $\{+i,-i\}$ & $\{+i,-i\}$ & $\{+i,-i\}$\tabularnewline
\hline 
$k_{z}=\pi$ & $-1$ & $\{+i,+i\}$ & $\{-i,-i\}$ & $\{-i,-i\}$ & $\{+i,+i\}$\tabularnewline
\hline 
\end{tabular}

\caption{Tables showing the channel eigenvalues and channel Chern number of
the two channels $I$ and $II$ for the hybrid-order topological insulator
with a translational-symmetry perturbation in the $x$-direction at
$k_{z}=0,\pi$.\label{tab:3}}
\end{table}

\end{document}